\documentstyle[epsfig,prl,twocolumn,aps,floats]{revtex} 
\preprint{T99/003}
\begin{document} 
\twocolumn[\hsize\textwidth\columnwidth\hsize\csname@twocolumnfalse%
\endcsname
\draft
\title{Edge logarithmic corrections probed by impurity NMR}
\author{V. Brunel, M. Bocquet and Th. Jolic\oe ur\thanks{e-mail~: 
vbrunel, thierry@spht.saclay.cea.fr}}
\address{Service de Physique Th\'eorique, CEA Saclay,
F91191 Gif-sur-Yvette, France} 
\date{\today} 
\maketitle
%%%%%%%%%%%%%%%%%%%%%%%%%%%%%%%%%%%%%%%%%%%%%%%%%%%%%%%%%%%%%%%%%
\begin{abstract} 
Semi-infinite quantum spin chains  display spin autocorrelations 
near the boundary with power-law exponents that are given by 
boundary conformal field theories. We show that NMR measurements on 
spinless impurities that break a quantum spin chain 
lead to a spin-lattice relaxation rate $1/T_{1}^{edge}$ that
has a temperature dependence which is a direct probe of the
anomalous boundary exponents. For the antiferromagnetic S=1/2
spin chain, we show that $1/T_{1}^{edge}\propto T \log^{2} T$
instead of $\log^{1/2} T$ for a bulk measurement.
We show that, in the case of a one-dimensional conductor described
by a Luttinger liquid, a similar measurement leads to a
relaxation rate $1/T_{1}^{edge}\propto T$,
independent of the anomalous exponent $K_{\rho}$.

\end{abstract}
%%%%%%%%%%%%%%%%%%%%%%%%%%%%%%%%%%%%%%%%%%%%%%%%%%%%%%%%%%%%%%%%%
] 

Impurity effects in quantum spin chains are a subject of
much interest since they are a sensitive probe
of the peculiar correlations that develop in these 
magnetic systems. For example it has been shown recently\cite{taki1} 
that an external uniform magnetic field creates a staggered
magnetization near impurities in the antiferromagnetic (AF) spin
chain compound Sr$_{2}$CuO$_{3}$, in agreement with theoretical
predictions\cite{EI1}. In S=1 AF spin systems with a Haldane gap 
in the bulk there are effective free spins S=1/2 at the 
end\cite{Hagiwara} of the chain that are revealed by the coupling 
to an impurity.

One-dimensional quantum systems like the S=1/2 AF chain
or the electron gas - the ``Luttinger liquid'' - are
critical systems with an effective low-energy physics
which is a conformal field theory. In both cases it is
the massless Gaussian model that is at the heart of 
the correct description. When  boundaries are introduced
in such problems, then it has been shown by Cardy that
there are boundary operators with nontrivial scaling 
dimensions\cite{cardy}. This means that if we measure
the spin autocorrelation in time 
$\langle {\bf S}(t)\cdot {\bf S}(0)\rangle$ for the
spin at the end of a chain, it decays with
a power law which is different from that of the bulk.
Measurements with a local probe like the $1/T_{1}$ relaxation rate
in NMR experiments are sensitive to this power law.

Conventional NMR is performed on nuclear spins that are
coupled by hyperfine interactions to the atomic
spin of the same atom~: for example in the S=1/2 antiferromagnetic 
spin chain Sr$_{2}$CuO$_{3}$ one may use $^{63}$Cu resonance\cite{taki1}. 
While this
is a local measurement, it involves the time autocorrelation 
of a {\it bulk} spin. We show in this Letter that by
NMR measurements on nuclei belonging to {\it spinless}
impurities {\it that break the chain} (or the 1D conductor)
one measures the anomalous boundary exponents of the boundary
conformal field theory. We call this ``edge NMR''.
In the case of the spin-1/2 chain we show that the bulk marginal
operator manifests itself by a logarithmic factor in $1/T_{1}$
which is different from the bulk case.

In the case of a spin chain, our main idea is illustrated in fig.~1.
Impurities  that are spinless but 
do have a nuclear spin are introduced
at a low concentration level in the spin chain.  The nuclear 
spin of the impurity in fig.~1  has substantial coupling only
with its two neighboring spins if we are dealing with a magnetic 
insulator. The coupling may be dipolar or transferred hyperfine~:
its precise nature does not matter for our reasoning.
When impurities are very dilute, the nuclear spin is surrounded
by two essentially semi-infinite spin chains.
So measurement of the relaxation rate $T_{1}$  of the impurity nuclear 
spin
involve {\it only} the fluctuations of the
two end spins and these spins give the same
contribution. If there is no exchange through the impurity,
then we have exactly a probe of a boundary critical phenomena since the
relaxation rate is given by the following formula\cite{Moriya}~:
\begin{equation} 
1/T_{1}=A^{2}\lim_{\omega\rightarrow 0}
\int^{+\infty}_{-\infty} dt\,\, e^{i\omega t}
\langle
{\bf S}_{0}(t)\cdot {\bf S}_{0}(0)
\rangle_{T}.
\label{T1} 
\end{equation} 
In this equation $A$ is the (hyperfine) coupling, ${\bf S}_{0}$
is the end spin of the semi-infinite chain and 
$\langle\ldots\rangle_{T}$ stands for thermal averaging.
An example would be
a S=1/2 spin chain realized by Cu$^{2+}$ ions and the impurity
may be Zn which is a possible substitution for Cu in the cuprate
family of materials. So Zn doping effectively cuts the chains.
The isotope $^{67}$Zn has a nuclear spin 
so NMR may be performed on this nucleus selectively.

In the case of the S=1/2 AF chain we show that 
$(1/T_{1})^{edge}\propto T \log^{2} T$
is the universal low-temperature behavior
instead of $\log^{1/2} T$ as predicted\cite{sachdev,sandvik}
and measured in the case of {\it bulk} NMR\cite{taki2}. In the case
of a one-dimensional conductor, one can perform the same kind
of measurement by using an impurity which is insulating,
i.e. which breaks the chain. Similarly one has now
a relaxation rate that scales as 
$(1/T_{1})^{edge}\propto T$ 
while in the bulk it is given by\cite{jerome}
$(1/T_{1})^{bulk}\propto T + T^{K_{\rho}}$ where $K_{\rho}$
is the anomalous exponent that characterizes the Luttinger liquid behavior
of the system.
\begin{figure}
\begin{center}
\mbox{\epsfig{file=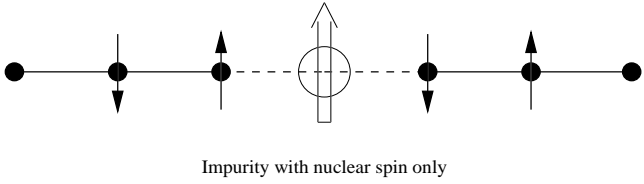,width=3.35in,angle=270}}
\end{center}
\caption{An	NMR	$1/T_{1}$ measurement performed	on an impurity
with a nuclear spin but no	atomic spin	(white dot) is a probe of 
the time autocorrelation of the end	spins of the spin chain (black dots).}
\label{f1}
\end{figure}

We first discuss the S=1/2 AF spin chain described by the
Heisenberg hamiltonian~:
\begin{equation}
	{\mathcal H}= J\sum_{n} S^{x}_{n}S^{x}_{n+1}
	+S^{y}_{n}S^{y}_{n+1}
	+\Delta S^{z}_{n}S^{z}_{n+1}.
	\label{Heis}
\end{equation}
In this equation, we have introduced the exchange anisotropy
$\Delta$ which is convenient for the bosonization description
of the low-energy physics. In most materials that are known
to be reasonably well described by a S=1/2 AF chain like
KCuF$_{3}$ or Sr$_{2}$CuO$_{3}$, in fact there is isotropy, i. e. 
$\Delta=1$. If $\Delta=0$ we have the XY chain which
maps onto a system of free spinless fermions in a band
of dispersion $E(k)=-J\cos k$ and two Fermi points 
at $k_{F}=\pm \pi /2$. There are particle-hole excitations
with small momentum transfer, i.e. staying on the same side
of the Fermi ``surface'' and also excitations across the Fermi
surface with momentum transfer $2k_{F} =\pi$. In the spin language,
they correspond respectively to ferro and antiferromagnetic 
fluctuations. In the bosonic language\cite{boso} we have~:
\begin{equation}
	S^{z}_{x}=-{1\over \sqrt{\pi}}\nabla_{x} \phi (x) 
	+{(-)^{x}\over \pi\alpha}
	\cos (\sqrt{4\pi} \phi (x)).
	\label{Sb}
\end{equation}
The effective theory in the infrared limit for the boson
$\phi$ is a free theory given by the following Lagrangian density~:
\begin{equation}
	{\mathcal L}_{\phi}={1\over 2}\left[(\partial_{t}\phi)^{2}-
	(\partial_{x}\phi)^{2}\right].
	\label{Hphi}
\end{equation}
For anisotropies $\Delta \leq 1$, the chain is in a massless
phase which is still described by Eq.(\ref{Hphi}) after a 
rescaling of the boson $\phi\rightarrow\phi/\sqrt{K}$
where $K=K(\Delta)$ is a parameter that contains the effects
of interactions induced by z-exchange. At the isotropic point
$\Delta =1$ we have $K=1/2$.

This massless effective theory leads to power laws
in the spin fluctuations. The local correlations
of a spin $\bf S$ in an infinite chain can be computed at nonzero
temperature  $T$ with the result~:
\begin{equation}
	\langle S^{z} (t) S^{z} (0)\rangle
	\propto {T^{2} \over \sinh ^{2} \pi Tt}
	        +({T^{2} \over \sinh^{2}\pi Tt})^{K},
	\label{sbz}
\end{equation}
\begin{equation}
		\langle S^{+} (t) S^{-} (0)\rangle
	\propto ({T^{2} \over \sinh^{2}\pi Tt})^{1/4K}
	     + ({T \over \sinh\pi Tt})^{2K+1/2K},
	\label{sb+}
\end{equation}
(omitting constant prefactors).
In the isotropic case both spin correlations are equal to
$T/\sinh \pi Tt$ and this behavior leads to a relaxation
$1/T_{1}$ calculated from Eq.~(\ref{T1}) which is $\propto T^{0}$.
This is essentially the behavior observed by Takigawa et al.\cite{taki2}.
There is an additional logarithmic correction $\log^{1/2} T$ due to
the presence of Umklapp processes for the fermions\cite{sachdev}.

If we now consider a spin at the end of an open chain, we
must supply our bosonic description with a boundary condition.
It is a Dirichlet condition which is appropriate\cite{EI2}.
The condition $\langle S^{z}(x=0)\rangle =0$ in the bosonic theory leads
to $\phi(0)=\sqrt{\pi}/2\sqrt{2}$ by use of Eq.~(\ref{Sb}). 
This boundary condition
can then be used to relate the chiral components $\phi_{L}$ and $\phi_{R}$.
Indeed $\phi_{R}$ may be regarded as the analytic continuation
of $\phi_{L}$ on the negative axis by
$\phi_{R}(x, t)=-\phi_{L}(-x, t)+\sqrt{\pi}/2\sqrt{2}$.
The Green's function can then be expressed in terms of the L field 
only and this field does not feel the existence of the boundary.

The edge autocorrelations have now a different behavior from the bulk.
We find that the uniform and staggered part of the correlation functions
have the same behavior at the edge, given by~:
\begin{equation}
	\langle S^{z} (t) S^{z} (0)\rangle
	\propto {T^{2} \over \sinh^{2}\pi Tt}
	\label{sez}
\end{equation}
\begin{equation}
		\langle S^{+} (t) S^{-} (0)\rangle
	\propto ({T \over \sinh\pi Tt})^{1/K},
	\label{se+}
\end{equation}
where we have kept the leading behavior. This is due to the fact that
spin operators have boundary scaling dimensions 1 instead of 1/2 in 
the bulk.
In the isotropic case
these correlations leads to contributions to $1/T_{1}$ that
scale as $T$ instead of $T^{0}$ in the bulk. So the boundary critical
behavior influences directly the NMR relaxation rate.
The leading behavior is thus $(1/T_{1})^{edge}\propto T$.

We now turn to the influence of the marginal operator.
Indeed it is known that there are logarithmic corrections to the power-law
behavior of the spin-1/2 chain\cite{logIan,log2}.
In fact the effective theory given by Eq.~(\ref{Hphi}) contains
an additional {\it marginal} operator that is allowed in the
isotropic case~: it is due to Umklapp scattering of the fermions and
can be written as $\cos(\sqrt{16\pi}\phi)$. The complete effective 
theory is given by~:
\begin{equation}
		{\mathcal L}_{\phi}={1\over 2}\left[(\partial_{t}\phi)^{2}-
	(\partial_{x}\phi)^{2}\right] + g \, \cos(\sqrt{16\pi}\phi).
	\label{margi}
\end{equation}
The set of the two coupling constants $K$ and $g$ flows
under renormalization group according to the Kosterlitz-Thouless
equations. There is a single line that reaches the fixed point
$K^{*}=1/2$ and which corresponds to an isotropic spin chain. Along this 
line, the flow of the couplings is marginal.
This operator leads
to a multiplicative logarithmic correction to the bulk magnetic 
susceptibility, as predicted theoretically\cite{EIAT} and observed 
experimentally\cite{MEU}. In the relaxation rate it leads to
a correction\cite{sachdev} $\propto (\log T)^{1/2}$ which is a 
possible explanation for the experimental upturn of $1/T_{1}$ at low 
temperature\cite{taki2}.
Many numerical studies have been devoted to demonstrate {\it ab initio}
its existence in the spin-1/2 chain. In fact  it is responsible
for the extremely slow approach to the thermodynamic limit\cite{Num,Barzy}.

This bulk marginal operator still exists in the theory with a 
boundary. In addition, some new operators are permitted
at the edge.
Note that the surface term $\partial_{x}\phi_{L}$,  while forbidden 
in the bulk of the chain, is a perfectly allowed operator in the 
boundary theory.
If we now classify all possible
marginal operators at the {\it boundary} fixed point, there are only
two such operators $\partial_{x}\phi_{L}$ and 
$\exp (i \sqrt{8\pi}\phi_{L})$. They have boundary scale dimension 1
and thus are marginal on the edge. 

But it is easy to see that
they are forbidden by the symmetry of the problem.
The rotational invariance around z-axis
forbids $\exp (i \sqrt{8\pi}\phi_{L})$ whereas it is
rotation around x-axis that forbids $\partial_{x}\phi_{L}$.
As a consequence, there are no extra marginal operators.
Thus the Green's function defined by~:
\begin{equation}
	G(t)^{edge}=	\langle S^{+}_{0} (t) S^{-}_{0} (0)\rangle ,
	\label{defG}
\end{equation}
obeys the following edge renormalization group equation~:
\begin{equation}
	\left[t{\partial\over\partial t} +\beta_{bulk} (g)
	{\partial\over\partial g}
	+2\gamma_{edge} (g) \right] G(t, g)^{edge} =0.
	\label{CSEq}
\end{equation}
In this equation, $g$ is the coupling of the marginal operator.
Since it is a bulk operator its $\beta$-function is that of the bulk~:
$\beta (g)\equiv \beta_{bulk}(g)$.
The anomalous dimension $\gamma_{edge}(g)$ however is related to
the short-distance singularity of the bulk marginal operator with
an operator that lies at the boundary. Hence it is modified
with respect to its bulk value.

We compute $\gamma_{edge}(g)$ by a Coulomb gas technique similar
to that used by Singh, Fisher and Shankar\cite{log2}, expressing
all the Coulomb gas quantities in terms of $\phi_{L}$ only.
We find~:
\begin{equation}
	\beta_{bulk}(g)= -2\pi g^{2} +O(g^{3}),
	\label{beta}
	\end{equation}
\begin{equation}
	\gamma_{edge}(g) = 1-2\pi g +O(g^{2}).
	\label{gamma}
\end{equation}
The result for $\gamma_{edge}$ is different from the bulk result
$\gamma (g) =1/2 -\pi g/2$~: the constant term is due to the edge scaling
dimension of the spin operator which is twice the bulk value. The $O(g)$
coefficient leads directly to the logarithmic correction.
The solution of Eq.~(\ref{CSEq}) by the method of characteristics
 leads to an asymptotic behavior
which includes a logarithmic correction~:
\begin{equation}
G(t)^{edge}\propto {1\over t^{2}}(\log t)^{2}.
\label{G1}
\end{equation}
This is the peculiar law of time decay for the autocorrelation
of the end spin of the chain (in the bulk it is $\log^{1/2}t/t$).
This result holds for the staggered as well as the uniform part
of the correlation function. We have checked that 
Eqs.~(\ref{beta},\ref{gamma}) are consistent with isotropy~:
$\langle S^{+}_{0} (t) S^{-}_{0} (0)\rangle$ and
$\langle S^{z}_{0} (t) S^{z}_{0} (0)\rangle$
have the same behavior.
To treat the effect of a nonzero temperature we impose periodic
conditions in Euclidean time with period $1/T$. The system lives then
on a half-cylinder. The Callan-Symanzik equation (\ref{CSEq})
is unchanged, though the Green's function depends upon an additional 
argument which is the temporal extent $1/T$. A standard finite-size
scaling argument leads then to a solution of the form~:
\begin{equation}
	G(t, T)^{edge}\propto 
	{T^{2}\over\sinh^{2}(\pi Tt)}(\log t)^{2}.
	\label{GT}
\end{equation}
Such a behavior leads to a relaxation rate 
$(1/T_{1})^{edge}\propto T\log^{2} T$
strikingly different from the bulk.
Experimental check of this result seems feasible with the possible 
complication that if the nearest-neighbor coupling needed in
edge NMR is dipolar then it may be that there are parasitic effects due
to the other neighboring chains.
We note that, while we find the well-known doubling
of surface exponents for the leading power law, this doubling
does not extend to the power of the logarithm.

{\it The Luttinger liquid.}
We now apply the same line of arguments to the one-dimensional
Luttinger liquid. There are many examples of physical systems
including quantum wires, carbon nanotubes or one dimensional
organic conductors that belong to the universality class
of the ``Luttinger liquid''. In such systems, the spin and charge
degrees of freedom are decoupled in the low-energy long-wavelength 
limit
and can be described in a bosonization
framework by two bosonic fields, $\phi_{\sigma}$ for the spin
and $\phi_{\rho}$ for the charge. For isotropic systems, the effective
theory for $\phi_{\sigma}$ is a free field theory like Eq.~(\ref{Hphi}).
The theory for $\phi_{\rho}$ is also a free theory but with a 
nontrivial constant $K_{\rho}$ which encapsulates all low-energy
long-distance properties of the system.
The rate $1/T_{1}$ is a probe of the spin fluctuations
through the spin density wave (SDW) correlations
$\langle SDW^{z}(t)\, SDW^{z}(0)\rangle$ where the $SDW^{z}$
operator is given by 
$c^{\dag}_{n\alpha}\sigma^{z}_{\alpha\beta} c_{n\beta}$
(we consider the isotropic case). In a translationally invariant 
system these correlations have a uniform part with $q\approx 0$
and a $q\approx 2k_{F}$ part. They both contribute to $(1/T_{1})^{bulk}$
leading to a law $T + T^{K_{\rho}}$ which involves explicitly
the coupling $K_{\rho}$. In the case of edge NMR, we use the bosonized
form of the SDW operator~:
\begin{equation}
	SDW^{z}_{staggered}\sim \sin \sqrt{2\pi K_{\rho}}\phi_{\rho}
	\, \sin \sqrt{2\pi}\phi_{\sigma} .
	\label{sdw}
\end{equation}
The condition $\langle S^{z}\rangle =0$ on the last spin of the chain
leads to the Dirichlet boundary conditions $\phi_{\sigma}(0)=0$
and $\phi_{\rho}(0)=\sqrt{\pi}/2\sqrt{2K_{\rho}}$.
The staggered part contributes then to $1/T_{1}$ through the product~:
\begin{eqnarray}
	&\langle \sin \sqrt{2\pi}   \phi_{\sigma}(t)
	\sin \sqrt{2\pi}  \phi_{\sigma}(0)\rangle_{\phi_{\sigma}(0)=0}
	\nonumber \\
	\times &\langle \sin \sqrt{2\pi K_{\rho}}  \phi_{\rho}(t)
	\sin \sqrt{2\pi K_{\rho}}  \phi_{\rho}(0) 
	\rangle_{\phi_{\rho}(0)={\sqrt{\pi}\over 2\sqrt{2K_{\rho}}}}.
	\label{prod}
\end{eqnarray}
The spin part in Eq.~(\ref{prod}) is exactly the same as in a
quantum spin chain while the charge part is different due to the
presence of $K_{\rho}$ but also due to the peculiar
boundary condition. As a consequence, when we compute these correlations
each factor involves four nonzero contributions instead of two in the 
bulk (the electric neutrality of Coulomb gas calculations is 
affected by the boundary). 
Let us compute the charge correlator between two arbitrary points
$(x_{1}, t_{1})$ and $(x_{2}, t_{2})$.
We give some details in the zero temperature case, a conformal mapping
can be used afterwards to obtain the corresponding expressions valid
for low $T$.
The left and right components
of the charge Boson are related by
$\phi_{\rho R}(x, t)=-\phi_{\rho L}(-x, t)+
{\sqrt{\pi}/ 2\sqrt{2K_{\rho}}}$ and the charge part $C_{\rho}$ 
of the correlator
Eq.~(\ref{prod}) is thus equal to~:
\begin{eqnarray}
	&\langle \cos \sqrt{2\pi K_{\rho}}  
	(\phi_{\rho L}(x_{1}, t_{1}) -\phi_{\rho L}(-x_{1}, t_{1}))
	\times
	\nonumber \\
	&\times \cos \sqrt{2\pi K_{\rho}}  
	(\phi_{\rho L}(x_{2}, t_{2}) -\phi_{\rho L}(-x_{2}, t_{2}))
	\rangle.
	\label{charge}
\end{eqnarray}
Evaluation of this correlation function leads to the following 
expression~:
\begin{equation}
	C_{\rho} = ({1\over x_{1}x_{2}})^{K_{\rho}/2}(I+{1\over I}),
	\label{C}
\end{equation}
where I is equal to~:
\begin{equation}
	\left[  (t_{1}-t_{2})^{2}-(x_{1}+x_{2})^{2}
	\over 
	(t_{1}-t_{2})^{2}-(x_{1}-x_{2})^{2}
	\right]^{K_{\rho}/2}.
	\label{Integ}
\end{equation}
When approaching the edge of the system, i.e. $x_{1}$ and
$x_{1}$ are $O(a)$ where $a$ is the lattice spacing, the expression
I of Eq.~(\ref{Integ}) goes to a constant and the correlator C
becomes independent of time
and the charge part in Eq.~(\ref{prod})
does not contribute to the time decay. 
This is to be contrasted with the spin correlator in Eq.~(\ref{prod}).
There the boundary condition is simply $\phi_{\sigma}(0)=0$ and
thus in the expression corresponding to Eq.~(\ref{C}), we find the
{\it difference} $I-1/I$ (with now $K_{\rho}$ replaced by 1)
that decays in time according to a power law $\sim 1/t^{2}$
We thus obtain a 
law of the form $T^{2}/\sinh^{2}\pi Tt$ at nonzero temperature
leading to a ``superuniversal'' behavior $1/T_{1}^{edge}\propto T$
independent of $K_{\rho}$.

We note that it has also been proposed recently\cite{EMK,EJM,MEJ}
to use photoemission spectroscopy on the open end of a Luttinger
liquid to probe  the boundary effects. In this case the physical 
measurement is sensitive to the boundary spectral properties
of the Luttinger liquid and also involves novel exponents
that differ from those of the bulk.

In this Letter, we have shown how edge NMR can be used as a 
probe of some boundary conformal field theories. For the S=1/2
AF spin chain, the relaxation rate scales like $T\log^{2}T$ at the edge
instead of $T^{0}\log^{1/2}T$. For a Luttinger liquid with gapless
spin and charge degrees of freedom it scales as $T$. This 
universal behavior is independent of the correlation 
exponent $K_{\rho}$ which rules all 1D behavior through anomalous  
power laws.

\acknowledgments

We thank I. Affleck for a useful correspondence and
F. David, J. F. Jacquinot, J. P. Renard for useful discussions 

%%%%%%%%%%%%%%%%%%%%%%%%%%%%%%%%%%%%%%%%%%%%%%%%%%%%%%%%%%%%%%%%%
%%%%%%%%%%%%%%%%%%%%%%%%%REFERENCES%%%%%%%%%%%%%%%%%%%%%%%%%%%%%%


\begin{references}

\bibitem{taki1}
M. Takigawa, N. Motoyama, H. Eisaki and S. Uchida,
Phys. Rev. B{\bf 55}, 14129 (1997).

\bibitem{EI1}
S. Eggert and I. Affleck,
Phys. Rev. Lett. {\bf 75}, 934 (1995).

\bibitem{Hagiwara}
M. Hagiwara {et al.}, 
Phys. Rev. Lett. {\bf 5}, 3181 (1990).

\bibitem{cardy}
J. Cardy, Nucl. Phys. B{\bf 240}, 514 (1984),
{\it ibid.}, B{\bf 270}, 186 (1986),
{\it ibid.}, B{\bf 275}, 200 (1986).

\bibitem{Moriya}
T. Moriya,
Prog. Theor. Phys. {\bf 28}, 371 (1962).

\bibitem{sachdev}
S. Sachdev,
Phys. Rev. B{\bf 50}, 13006 (1994).

\bibitem{sandvik}
A. W. Sandvik,
Phys. Rev. B{\bf 52}, 9831 (1995).

\bibitem{taki2}
M. Takigawa, N. Motoyama, H. Eisaki and S. Uchida,
Phys. Rev. Lett. {\bf 76}, 4612 (1996).

\bibitem{jerome}
See e. g. the contribution by
D. J\'erome, in {\it Organic Superconductors: from (TMTSF)$_{2}$PF$_{6}$
to Fullerenes}, p. 405, M. Dekker, New York, 1994.

\bibitem{boso}
see e.g. papers in M. Stone (ed.), {\it Bosonization},
World Scientific, Singapore, 1994.

\bibitem{EI2}
S. Eggert and I. Affleck,
Phys. Rev. B{\bf 46}, 10866 (1992).

\bibitem{logIan}
I. Affleck, D. Gepner, H. J. Schulz and T. Ziman,
J. Phys. A{\bf 22}, 511 (1989).

\bibitem{log2}
R. R. P. Singh, M. Fisher and R. Shankar,
Phys. Rev. B{\bf 39}, 2562 (1989).

\bibitem{EIAT}
S. Eggert, I. Affleck and M. Takahashi,
Phys. Rev. Lett. {\bf 73}, 332 (1994).

\bibitem{MEU}
N. Motoyama, H. Eisaki and S. Uchida,
Phys. Rev. Lett. {\bf 76}, 3212 (1996).

\bibitem{Num}
K. Hallberg, P. Horsch and G. Martinez,
Phys. Rev. B{\bf 52}, R719 (1995).
T. Koma and N. Mizukoshi,
J. Stat. Phys. {\bf 83}, 661 (1996).

\bibitem{Barzy}
V. Barzykhin and I. Affleck,
e-print cond-mat/9810075.

\bibitem{EMK}
S. Eggert, A. Mattsson and J. M. Kinarets,
Phys. Rev. B{\bf 56}, 15537 (1997).

\bibitem{EJM}
S. Eggert, H. Johannesson and A. Mattsson,
Phys. Rev. Lett. {\bf 76}, 1505 (1996).

\bibitem{MEJ}
A. Mattsson, S. Eggert and H. Johannesson,
Phys. Rev. B{\bf 56}, 15615 (1997).


\end{references}
\end{document}